\documentclass[journal=ancac3,manuscript=article]{achemso}
\usepackage{amsmath,amssymb}
\usepackage{graphicx}
\usepackage{siunitx}
\usepackage{upgreek}
\usepackage[final]{changes}
\title{\replaced{Chip}{Wafer}-Scale Aligned Chiral Carbon Nanotubes Exhibiting Giant Second Harmonic Generation}

\author{Rui Xu}
\altaffiliation{These authors contributed equally to this work.}
\affiliation{Department of Materials Science and NanoEngineering, Rice University, Houston, TX 77005, USA}

\author{Jacques Doumani}
\altaffiliation{These authors contributed equally to this work.}
\affiliation{Department of Electrical and Computer Engineering, Rice University, Houston, TX 77005, USA}

\author{Viktor Labuntsov}
\affiliation{Department of Electrical Engineering, University at Buffalo, Buffalo, NY 14228, USA}

\author{Nina Hong}
\affiliation{J.A. Woollam Co., Inc., Lincoln, NE 68508, USA}

\author{Anna-Christina Samaha}
\affiliation{Lebanese University, Jdeidet 90656, Lebanon \\
Laboratory of Biomaterials and Intelligent Materials; Department of Physics}

\author{Weiran Tu}
\affiliation{Department of Materials Science and NanoEngineering, Rice University, Houston, TX 77005, USA}

\author{Fuyang Tay}
\affiliation{Department of Electrical and Computer Engineering, Rice University, Houston, TX 77005, USA}

\author{Elizabeth Blackert}
\affiliation{Department of Materials Science and NanoEngineering, Rice University, Houston, TX 77005, USA}

\author{Jiaming Luo}
\affiliation{Department of Materials Science and NanoEngineering, Rice University, Houston, TX 77005, USA}

\author{Mario El Tahchi}
\affiliation{Lebanese University, Jdeidet 90656, Lebanon \\
Laboratory of Biomaterials and Intelligent Materials; Department of Physics}

\author{Weilu Gao}
\affiliation{Department of Electrical and Computer Engineering, University of Utah, Salt Lake City, UT 84112, USA}

\author{Jun Lou}
\affiliation{Department of Materials Science and NanoEngineering, Rice University, Houston, TX 77005, USA}

\author{Yohei Yomogida}
\affiliation{Department of Physics, Tokyo Metropolitan University, Tokyo 192-0397, Japan}

\author{Kazuhiro Yanagi}
\affiliation{Department of Physics, Tokyo Metropolitan University, Tokyo 192-0397, Japan}

\author{Riichiro Saito}
\affiliation{Department of Physics, Tokyo Metropolitan University, Tokyo 192-0397, Japan}
\alsoaffiliation{Department of Physics, Tohoku University, Sendai 980-8578, Japan}
\alsoaffiliation{Department of Physics, National Taiwan Normal University, Taipei 106, Taiwan}

\author{Vasili Perebeinos}
\affiliation{Department of Electrical Engineering, University at Buffalo, Buffalo, NY 14228, USA}
\email{vasilipe@buffalo.edu}

\author{Andrey Baydin}
\affiliation{Department of Electrical and Computer Engineering, Rice University, Houston, TX 77005, USA}
\alsoaffiliation{Smalley-Curl Institute, Rice University, Houston, TX 77005, USA}
\email{baydin@rice.edu}

\author{Junichiro Kono}
\affiliation{Department of Materials Science and NanoEngineering, Rice University, Houston, TX 77005, USA}
\alsoaffiliation{Department of Electrical and Computer Engineering, Rice University, Houston, TX 77005, USA}
\alsoaffiliation{Smalley-Curl Institute, Rice University, Houston, TX 77005, USA}
\alsoaffiliation{Rice Advanced Materials Institute, Rice University, Houston, TX 77005, USA}
\email{kono@rice.edu}

\author{Hanyu Zhu}
\affiliation{Department of Materials Science and NanoEngineering, Rice University, Houston, TX 77005, USA}
\alsoaffiliation{Department of Electrical and Computer Engineering, Rice University, Houston, TX 77005, USA}
\alsoaffiliation{Smalley-Curl Institute, Rice University, Houston, TX 77005, USA}
\alsoaffiliation{Rice Advanced Materials Institute, Rice University, Houston, TX 77005, USA}
\email{hz67@rice.edu}


\begin{document}

\maketitle
\newpage
\begin{abstract}
Chiral carbon nanotubes (CNTs) are direct-gap semiconductors with optical properties governed by one-dimensional excitons with enormous oscillator strengths. Each species of chiral CNTs has an enantiomeric pair of left- and right-handed CNTs with nearly identical properties, but enantiomer-dependent phenomena can emerge, especially in nonlinear optical processes. Theoretical studies have predicted strong second-order nonlinearities in chiral CNTs, but no experimental quantitative verification has been reported due to the lack of macroscopically ordered assemblies of single-enantiomer chiral CNTs. Here, \deleted{for the first time, }we report the synthesis of centimeter-scale, densely packed, aligned single-enantiomer chiral CNT films that are microfabrication-compatible. We observe giant second harmonic generation (SHG) emission from the chiral CNT film, which originates from the intrinsic chirality and inversion symmetry breaking of the atomic structure of chiral CNTs. The observed \replaced{nonlinear susceptibility of the as-fabricated film reaches $4.9\times 10^2$\,pm/V at a pump wavelength of 1030\,nm, corresponding to the lowest-energy excitonic resonance, indicating $\chi_{xyz} = 1.6\times 10^3$\,pm/V for a perfectly aligned CNT crystal}{value of the dominant element of the second-order nonlinear optical susceptibility tensor reaches 1.5 $\times 10^3$\,pm/V at a pump wavelength of 1030\,nm, corresponding to the lowest-energy excitonic resonance}. Our calculations based on many-body theory correctly estimate the spectrum and magnitude of such excitonically enhanced optical nonlinearity.  
These results are promising for the development of scalable chiral-CNT electronics and nonlinear photonics.    
\end{abstract}

\section{Introduction}
Structural chirality is a fundamental property in which the material is not identical to its mirror image. Materials possessing chirality can enable intriguing properties, including optical activity~\cite{noauthor_circular_nodate,ahn_new_2017}, magneto-chiral effects~\cite{rikken_observation_1997,train_strong_2008}, and chiral-induced spin selectivity~\cite{naaman_chiral-induced_2012,yang_real-time_2023}, which can be useful for information processing and transfer applications~\cite{yang_circularly_2013,brandt_added_2017,yang_chiral_2021,wei_chiral_2021}. Due to lifted inversion symmetry in chiral materials, second-order optical nonlinearity is also expected for conventional chiral crystalline insulators including $\alpha$-quartz and TeO$_2$~\cite{singh_violation_1972,chemla_optical_2003} and recently discovered chiral semiconductors such as hybrid organic-inorganic halides~\cite{yuan_chiral_2018, yao_strong_2021, ge_chiral_2022}, with resonance-enhanced nonlinear optical susceptibilities $\chi^{(2)}$ of up to 100\,pm/V. However, it is counterintuitive that second harmonic generation (SHG) can arise purely from structural chirality without any local electrical or magnetic polarization, such as in non-magnetic elemental allotropes. Indeed, Kleinman symmetry dictates that SHG is forbidden in isotropic chiral materials without dispersions~\cite{boyd_nonlinear_1992}. Previously, crystalline anisotropy and large dispersions have enabled large SHG from tellurium, a chiral elemental allotrope~\cite{cheng_large_2019, londono-calderon_intrinsic_2021, wang_anisotropic_2021}, but the nonlinear susceptibility was only quantified at far-infrared wavelengths, instead of near the resonance at the bandgap~\cite{shermanAbsoluteMeasurementSecondharmonic1973}. 

Chiral carbon nanotubes (CNTs) are ideal candidates for exploring chirality-induced nonlinear optical phenomena. Since their discovery, CNTs have shown unique electronic, photonic, and mechanical properties arising from one-dimensional (1D) quantum confinement effects~\cite{AvourisEtAl2008NP,BlackburnEtAl2018AMb,Li2021AN,BaydinEtAl2022M,ZhuEtAl2023AM}. 
Calculations have predicted strong 1D van Hove singularity-enhanced second-order nonlinearities, $\chi^{(2)}$ on the order of nm/V, for chiral species of single-wall CNTs, where their chirality indices ($n$,$m$) satisfy $n \ne m$ and $m \ne 0$~\cite{guo_linear_2004,rerat_ab_2018}. Such chiral CNTs are semiconductors with diameter ($d_{\rm t}$)-dependent direct band gaps, $E_{\rm g}$\,(eV)~$\sim$~0.7/$d_{\rm t}$\,(nm), potentially useful for integrated nanophotonics and chiral optoelectronics.~\cite{weisman_introduction_2017} Being 1D semiconductors with $d_{\rm t} \sim 1$\,nm, chiral CNTs exhibit enormous excitonic effects, which can resonantly boost SHG processes further~\cite{nishihara_empirical_2022}. Several experiments have observed SHG from different types of CNT materials, including single tubes~\cite{huttunen_measurement_2013}, thin films~\cite{okawara_second-harmonic_2019,dominicis_symmetry_2004}, and chiral CNTs embedded in host porous crystals~\cite{su_resonant_2008}. However, the observed optical nonlinearity was very weak (effective $\chi^{(2)} \sim 0.01$\,pm/V) due to the coexistence of both enantiomers, random CNT orientations, and low packaging densities, hindering both fundamental studies and practical applications. Furthermore, the observed SHG may have originated from surface effects or local defects instead of the intrinsic chirality of CNTs since the symmetry of the observed signal disagreed with the expected SHG tensor of chiral CNTs, which should contain only two non-zero elements, i.e., $\chi^{(2)}_{xyz}$ and $\chi^{(2)}_{yzx}$ ($= -\chi^{(2)}_{xyz}$). \added{Such second order nonlinearity is only possible with intrinsic molecular-level chirality, in contrast to} \deleted{Recent efforts have demonstrated that wafer-scale CNT films with} engineered structural chirality \deleted{can exhibit strong optical activity and chiroptical responses} through twist-patterned or helically aligned assemblies \added{which can only show linear chiroptical responses}~\cite{doumani_engineering_2023,Kono_Chiralized,fan_programmable_2025}. 

Here we \deleted{utilize the intrinsic molecular chirality of single-enantiomer CNTs to} construct densely packed, aligned \added{centimeter-scale} thin films \added{of single-enantiomer CNTs} \deleted{that inherently break inversion symmetry without requiring geometric asymmetry or structural patterning. We report the fabrication of centimeter-scale films of aligned enantiomer-pure chiral CNTs and the observation of} to observe giant SHG from \replaced{their intrinsic chirality}{the films}. We chose (6,5)$^-$ CNTs for this study; (6,5)$^+$ and (6,5)$^-$ are the left-handed and right-handed enantiomers, respectively, of the otherwise identical CNT species with $d_{\rm t} = 0.76$\,nm and $E_{\rm g} \sim 1$\,eV. Incident angle- and polarization-dependent measurements of SHG, combined with analysis of the $\chi^{(2)}$ tensor, confirmed that the observed SHG signal originated from the intrinsic structural chirality of the (6,5)$^-$ CNTs. The excitation-wavelength-dependence of the SHG signal exhibited a prominent peak at around the $E_{11}$ exciton resonance ($\sim$1\,$\upmu$m), in agreement with the results of our calculations based on many-body theory. Excitation power dependence showed a deviation from the expected quadratic behavior for excitation intensities above a threshold of $\sim0.2$\,mJ/cm$^2$, which we attribute to absorption saturation due to exciton-exciton annihilation~\cite{murakami_nonlinear_2009}. From the measured SHG signals, we \added{derived the effective nonlinear susceptibility of the as-fabricated thin film to be $4.9\times 10^2$\,pm/V from experimentally obtained properties without any free parameters, and} estimated the value of $\chi^{(2)}_{xyz}$ to be $1.6\times 10^3$\,pm/V \added{for a perfect chiral CNT crystal}, which is comparable to the values reported for single-crystalline two-dimensional semiconductors and Weyl semimetals~\cite{janisch_extraordinary_2014, kumar_second_2013, you_nonlinear_2019, kim_three-dimensional_2024,wu_giant_2017}. 
These results demonstrate that aligned enantiomer-pure CNT films, which can be easily transferred to any substrate with excellent conformity and scalability, are promising for applications in \added{sub-wavelength nonlinear optics,} integrated chiral nano-photonics, and chiral quantum optics~\cite{lodahl_chiral_2017}. \added{Compared with other nonlinear nanomaterials, such as boron nitride nanotubes and NbOCl$_2$, chiral CNT exhibits an order of magnitude higher intrinsic nonlinear susceptibility in the telecom wavelengths for optical communication and silicon photonics applications~\cite{abdelwahab_giant_2022,ma_strong_2024}. CNTs may also be combined with other sub-wavelength nonlinear optical structures and engineering techniques, such as quasi-phase matching by periodic stacking oppositely handed films, dielectric nonlinear resonators, and metasurfaces to enhance their nonlinear performance~\cite{koshelev_subwavelength_2020, trovatello_quasi-phase-matched_2025, krasnok_nonlinear_2018, li_nonlinear_2017}.}

\section{Results and Discussion}

\subsection{Aligned Thin Film of Enantiomer-Pure Single-Chirality (6,5)$^-$ CNTs}

\begin{figure*}[t]
    \centering
    \includegraphics[width=\linewidth]{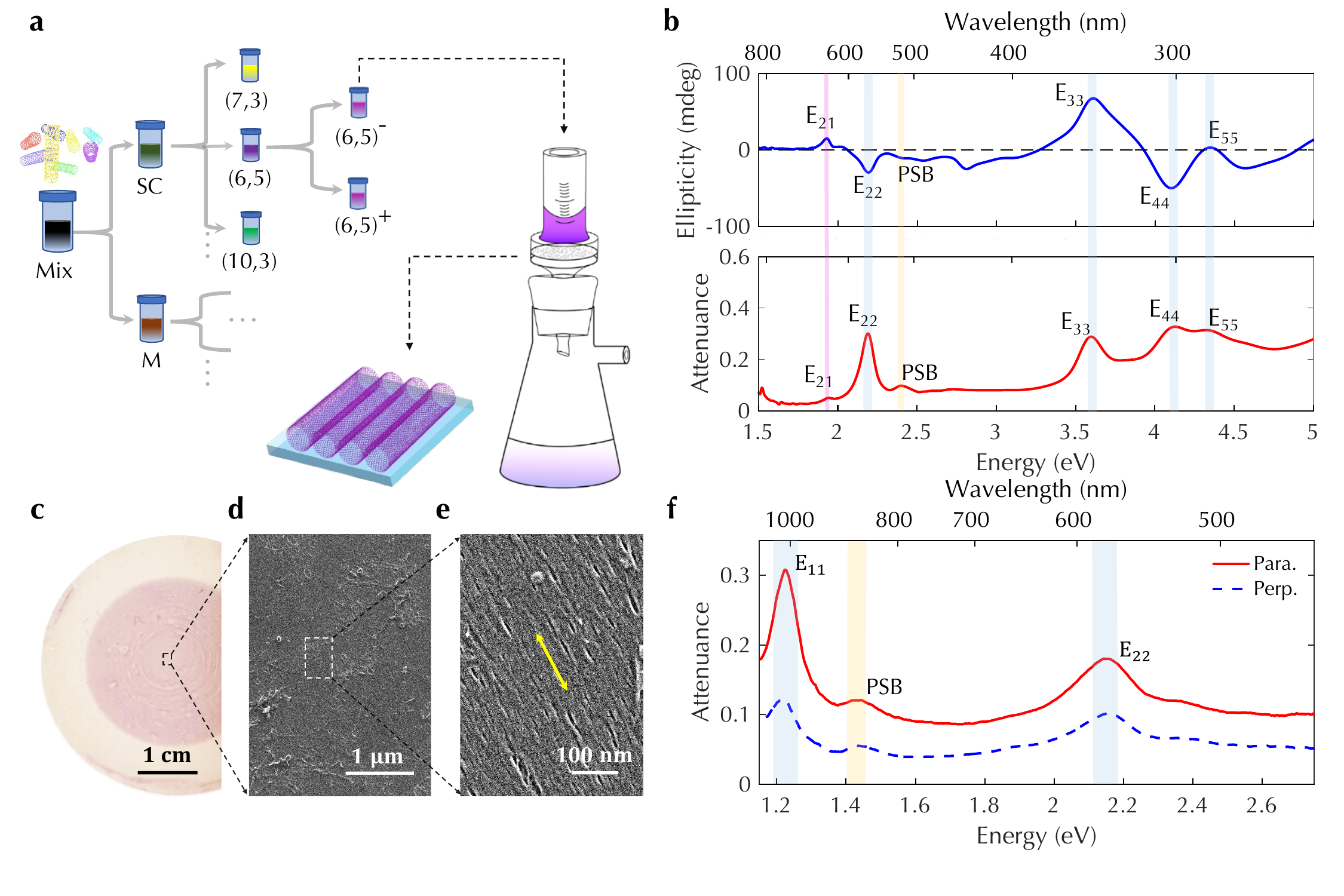}
    \caption{
        Fabrication and characterization of an aligned film of enantiomer-pure (6,5)\textsuperscript{--} CNTs. 
        \textbf{(a)} An as-purchased mixture of CNTs was dispersed in water and separated by metallicity and chirality via multi-step gel chromatography. A (6,5)\textsuperscript{--} CNT suspension was extracted, and an aligned film was fabricated using controlled vacuum filtration (CVF) and transferred onto a fused silica substrate. 
        \textbf{(b)} Circular dichroism and attenuance spectra for the (6,5)\textsuperscript{--} CNT suspension. Analysis yielded an enantiomeric purity of 91.6\%. 
        \textbf{(c)} Optical image of the chiral aligned (CA) CNT thin film with a diameter of 3~cm. 
        \textbf{(d,e)} SEM images of the CA CNT film at low and high magnifications, respectively, demonstrating its uniformity, high packing density, and microscopic in-plane anisotropy. Yellow arrow indicates the alignment direction. 
        \textbf{(f)} \replaced{Local}{Global} linearly polarized attenuance spectra for the CA sample with a probed area of $\sim$$1.4 \times 10^{3}$~µm$^2$. The observed peaks correspond to the $E_{11}$ exciton resonance, its phonon sideband (PSB), and the $E_{22}$ exciton. The linear dichroism value at $E_{11}$ is 0.9, corresponding to a nematic order parameter of $\sim$0.45.
    }
    \label{figure.1}
\end{figure*}

We first prepared an aqueous suspension of enantiomer-enriched (6,5)$^-$ CNTs and measured its enantiomer purity ($EP$). The suspension was purified using gel chromatography~\cite{yomogida_industrial-scale_2016,yomogida_automatic_2020,wei_high-yield_2018}, as illustrated in Fig.~\ref{figure.1}\textbf{a} and described in more detail in Methods.
The $EP$ of the suspension was characterized by circular dichroism (CD) spectroscopy; see Fig.~\ref{figure.1}\textbf{b}. The CD intensity spectrum exhibits multiple peaks corresponding to the $E_{ij}$ excitonic transitions shown in the attenuation spectrum; here, $i$ and $j$ are integers, and transitions with $i=j$ are strongly allowed for light polarization parallel to the nanotube axis~\cite{WeismanKono19Book}. The $EP$ of the (6,5)$^-$ CNT suspension can be calculated from the CD intensity (in mdeg) normalized by the attenuance $A$ at the $E_{22}$ transition, denoted by $CD_{E_{22}}$ and $A_{E_{22}}$, respectively, as follows~\cite{wei_determination_2017}:
\begin{align}
EP(6,5)^- = \left ( 50-0.421\frac{CD_{E_{22}}}{A_{E_{22}}} \right )\% = 91.6\%,
\label{eq:omebroad}
\end{align}
where we used $CD_{E_{22}} = -29.67$\,mdeg and $A_{E_{22}} = 0.30$; see Figure S1 for zoom-in data. This $EP$ value is comparable with the state-of-art in the literature~\cite{wei_length-dependent_2023}.

We then fabricated three CNT films using the CVF method~\cite{he_wafer-scale_2016} and transferred them onto fused silica substrates (Fig.~\ref{figure.1}\textbf{c} and Methods). The first film, labeled `chiral aligned' (CA), contained aligned enantiomer-pure (6,5)$^-$ CNTs. The film preserved the single-chirality nature of the original CNT suspension, as evidenced by Raman spectra (Supporting Information Section 1)~\cite{PhysRevB.72.205438}. The second sample, labeled `chiral random' or CR, contained randomly oriented enantiomer-pure (6,5)$^-$ CNTs prepared from the same suspension as the CA film. Finally, the third film, labeled `racemic aligned' or RA, consisted of highly aligned CNTs containing metallic and semiconducting CNTs with mixed chiralities~\cite{he_wafer-scale_2016}. All three films had the same thickness of around 20\,nm, measured by atomic force microscopy (Supporting Information Section 2). \added{The thickness of the film produced by CVF was shown to be homogeneous over large-area samples.}~\cite{doumani_engineering_2023} Scanning electron microscope images indicated that these films were densely packed with CNTs (Fig.~\ref{figure.1}\textbf{d} and Figure S3). \added{Although individual CNTs cannot be resolved by SEM due to their sub-nanometer diameters, the observed elongated nanoscale morphological features, corresponding to bundled CNTs, are consistent with the macroscopic alignment direction of the film. The sample has a nearly ideal local packing density on the order of $\sim~1$ CNT/nm$^2$ from transmission electron microscopy,~\cite{he_wafer-scale_2016} although gaps between aligned bundles can reduce the density by 8--12\% at 100-nm scale from perfect CNT crystals.~\cite{doumani_engineering_2023}}

Polarized attenuance spectra for the CA film (Fig.~\ref{figure.1}\textbf{f}) confirmed its global in-plane anisotropy within a probed area of $1.4\times10^{3}$\,$\upmu$m$^2$. The first three peaks in the attenuance spectra correspond to the $E_{11}$ exciton resonance (1.23\,eV), the phonon sideband of the $E_{11}$ exciton resonance (PSB, 1.43\,eV), and the $E_{22}$ exciton resonance (2.15\,eV), respectively. 
The difference in $E_{11}$ attenuance between parallel ($A_H$) and perpendicular ($A_V$) polarizations with respect to the CNT alignment direction yields a linear dichroism (LD) of $2 \times (A_{\rm H} - A_{\rm V})/(A_{\rm H} + A_{\rm V}) = 0.9$, corresponding to a nematic order parameter $S=0.45$ 
within the probed area~\cite{katsutani_direct_2019}. On a more local scale within an optically resolvable area of 1.3\,$\upmu$m$^2$, the CA film showed stronger alignment with $S=0.66\sim~0.71$. 
\added{We also characterized the nematic order parameter of the CA film over probe size as large as $2.2~\mathrm{mm}$, yielding $S \approx 0.24$, while still indicating finite alignment at the chip scale (Supporting Information Section 4).}

\begin{figure*}[t]
    \centering
    \includegraphics[width=\linewidth]{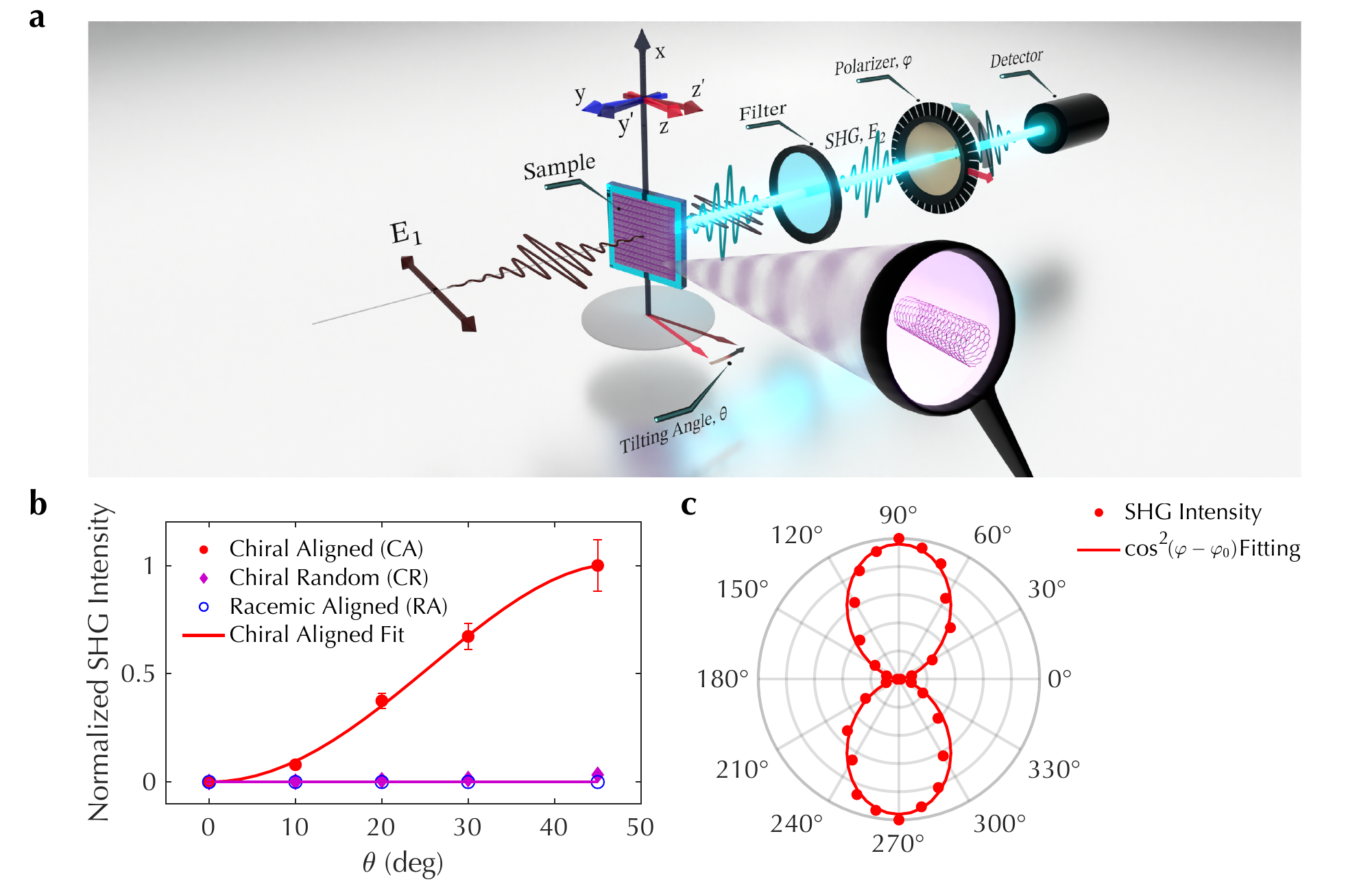}
    \caption{
        Quantifying the SHG tensor elements of aligned enantiomer-pure (6,5)\textsuperscript{--} CNT thin films. 
        \textbf{(a)} The excitation pulse is horizontally polarized ($z$-axis), with its electric field at an angle $\theta$ relative to the CNT alignment direction. The emitted SHG is separated from the excitation pulse by an optical filter and analyzed using a polarizer at an angle $\varphi$ from the excitation polarization. Axes with (without) the prime denote the lab (sample) coordinate system. Inset: the chiral atomic structure of a single (6,5)\textsuperscript{--}, a left-handed CNT. 
        \textbf{(b)} Strong SHG emission is observed from the chiral enantiomer-pure and aligned (CA) film at non-zero tilt angle, consistent with the $\chi_{yxz}^{(2)}$ tensor element, while negligible SHG is detected from the chiral random (CR) or racemic aligned (RA) films. 
        \textbf{(c)} The SHG radiation is vertically polarized, orthogonal to the excitation pulses, in agreement with SHG tensor symmetry. The fitting parameter $\varphi_0$ denotes the angle between SHG polarization and the $z$-axis in the lab frame.
    }
    \label{figure.2}
\end{figure*}

\subsection{Quantitative Second Harmonic Generation Spectroscopy}

Next, we resonantly pumped the CA film near the $E_{11}$ exciton transition energy ($\sim1.2$\,eV) and measured the emission at the second harmonic ($\sim2.4$\,eV). 
In chiral CNTs, such as (6,5) CNTs, only one independent nonzero element, $\chi_{x'y'z'}^{(2)} = -\chi_{y'x'z'}^{(2)}$, exists in the second order nonlinear optical susceptibility tensor~\cite{guo_linear_2004}, where $z'$ denotes the nanotube axis in the sample coordinate system (Fig.~\ref{figure.2}\textbf{a} and Figure S8). Therefore, the intensity of SHG, $I_{\rm SHG}$, arising from the intrinsic structural chirality should be nonzero only when the electric field of the incident light has finite projections onto both the $y'$ and $z'$ axes in the sample coordinate system:
\begin{align}
I_{\rm SHG} \propto P_{x'}^2= \left( 2\lvert \chi_{x'y'z',\text{eff}}^{(2)} \rvert E_{y'} E_{z'} \right)^2,
\label{eq:shg}
\end{align}
where $E_{y'}$ and $E_{z'}$ are the electric field projections of the fundamental field onto the $y'$ and $z'$ axes, respectively. The emission dipole $P_{x'}$ is expected to be linearly polarized along the $x'$ axis (Figure S8). As illustrated in Fig.~\ref{figure.2}a, the CNT film was initially aligned with the polarization of the fundamental beam along the $z$-axis in the laboratory coordinate system and then tilted around the $x$-axis by an angle $\theta$ (Methods). We observed strong SHG emission only from the CA film only when $\theta \ne 0^\circ$; under normal incidence ($\theta = 0^\circ$), no SHG was observed due to the absence of $E_{y'}$, as expected from Eq.~\ref{eq:shg}. The angular dependence of the SHG emission intensity can be fit well by Eq.\,S17, which includes the effects of refraction and phase matching. At $\theta=45^\circ$, we further confirmed that the polarization of the SHG radiation can be fit by $\cos^2(\varphi-\varphi_{0})$, where $\varphi_{0}$ is the angle between the SHG polarization and the $z$ axis in the laboratory system. As shown in Fig.~\ref{figure.2}\textbf{c}, we used $\varphi_{0}$ as a fitting parameter and obtained $\varphi_{0} = 87.5^\circ$ as the optimum value, indicating that the SHG emission is vertically polarized along the $x'$ ($x$) axis in the sample (laboratory) coordinate system, consistent with Eq.~\ref{eq:shg}. The combined tilting angle- and polarization-dependent measurements provided strong evidence that the observed SHG signal stems from the intrinsic chirality of the CNTs instead of surface symmetry breaking, which should yield $y'$-polarized emission, or local defects in the film, which should give a finite SHG signal under normal incidence. 
We observed small nonzero SHG from the CR film with an intensity less than 3\% of that from the CA film, likely due to a finite nematic order through local spontaneous alignment. \added{Such low SHG in SR also means the signal from CA cannot be attributed to extrinsic factors like surfactants.}
We observed no SHG emission from the RA film within the detection limit. These results highlight the essential role of both enantiomer purity and alignment for observing strong SHG emission from films of chiral CNTs.

 
\begin{figure*}[t]
    \centering
    \includegraphics[width=\linewidth]{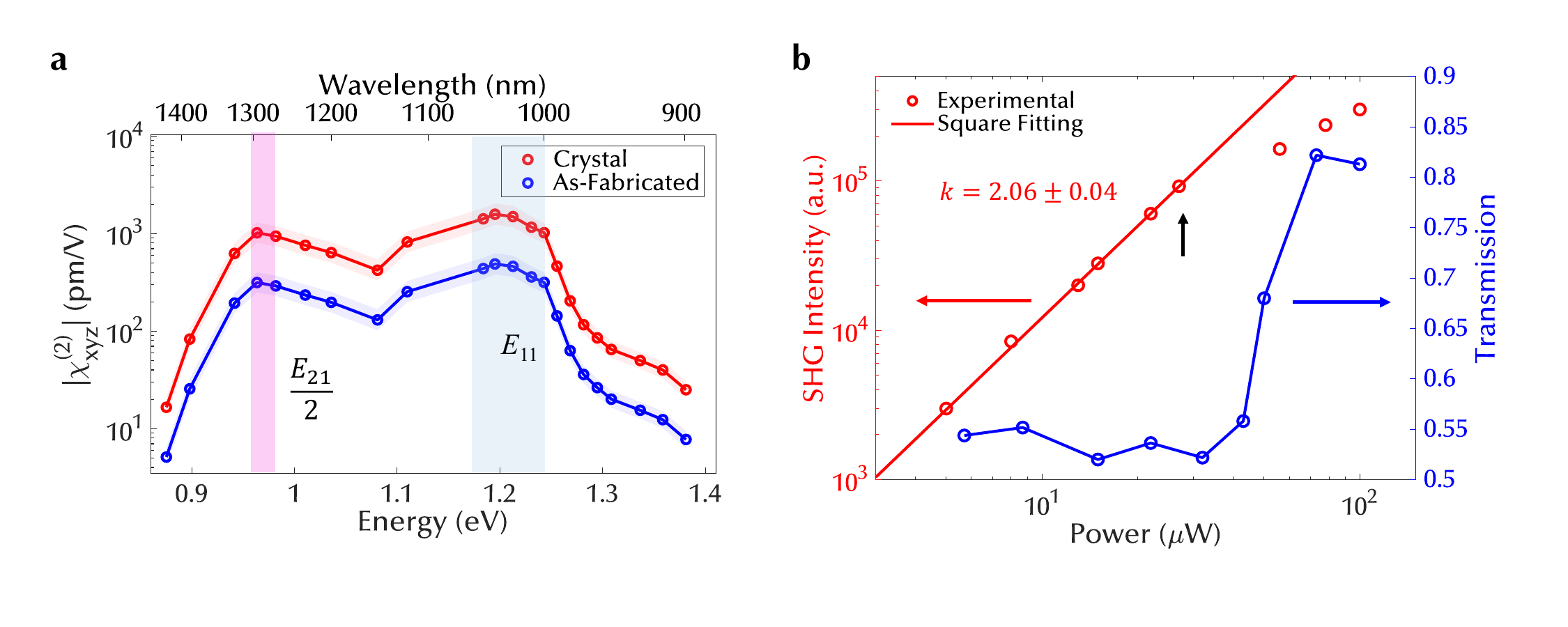}
    \caption{
        Photon-energy and fluence dependence of the observed second-order optical nonlinearity of the aligned enantiomer-pure (6,5)\textsuperscript{--} CNT. 
        \textbf{(a)} Experimentally determined effective nonlinear susceptibility of \added{the as-fabricated film (blue curve) and $\chi_{xyz}^{(2)}$ of a perfect CNT crystal after correcting the non-idealities of the sample (red curve).} The peak values \replaced{are obtained}{approaching $1.5\times10^3$~pm/V} for excitation photon energy at around 1.2~eV near the $E_{11}$ exciton resonance (blue bar) and an additional peak at around 1~eV near half of the $E_{12}$ exciton resonance (purple bar)\deleted{, after correcting for the effects of multiple reflections, phase mismatch, and the collection efficiency of the setup}. The width of the shaded area corresponds to the error range. 
        \textbf{(b)} Excitation fluence dependence of the SHG intensity (left axis) measured at a fundamental photon energy of 1.23~eV. The data show a deviation from a quadratic fit above the saturation absorption threshold of $\sim$0.2~mJ/cm$^2$, determined by power-dependent transmission measurements (right axis). The black arrow marks the excitation fluence used in panel \textbf{(a)}. Below this, SHG exhibits a power-law dependence with exponent $2.04\pm0.06$.
    }
    \label{figure.3}
\end{figure*}

We determined the value of $\lvert\chi_{xyz}^{(2)}\rvert$ from the SHG data for the CA film as a function of excitation photon energy in the range between 0.87\,eV and 1.37\,eV, finding strong excitonic enhancement (Figure~\ref{figure.3}\textbf{a}). \added{Here we distinguish $\chi_{xyz,\mathrm{eff}}^{(2)}$ of the as-synthesized film from the intrinsic second-order nonlinear susceptibility of a perfectly aligned, densely packed, enantiomerically pure CNT crystal.} To experimentally quantify $\lvert\chi_{xyz,\mathrm{eff}}^{(2)}\rvert$, we considered multiple reflections, phase mismatch, and the collection efficiency of the setup\added{, all determined independently from experiment}. \added{This effective nonlinearity of the film (blue curve in Fig. 3a) is corrected by the experimentally measured enantiomer purity (0.90), nematic order (0.45), filling factor considering the microscopic gaps between CNT bundles (0.88), and incomplete film coverage at the macroscopic scale (0.85) to yield the intrinsic nonlinearity of the CNT crystal (red curve, Supporting Information Section 7)}. The \added{resulting spectrum, obtained without any free fitting parameter}, shows a prominent peak centered at 1.21\,eV, closely matching the absorption resonance of the $E_{11}$ exciton observed in the linear optical susceptibility $\lvert\chi_{z}^{(1)}\rvert$ obtained through ellipsometry measurements (Supporting Information Section 6). 
The additional peak centered at 0.96\,eV is attributed to the emission resonance with $E_{12}$ exciton (Figure~\ref{figure.1}\textbf{b}), whose optical dipole is in the $xy$ plane instead (Supporting Information Section 8). The maximum measured $\chi_{xyz,\mathrm{eff}}^{(2)}$ is $4.9\times 10^2$\,pm/V, which corresponds to a $\chi_{xyz}^{(2)}$ of $1.6\times 10^3$\,pm/V in a perfect chiral CNT crystal. 
This is the largest experimentally verified value in any 1D material systems, to the best of our knowledge, and is comparable with the highest $\chi^{(2)}$ values in the near-infrared region reported for any other nonlinear optical materials~\cite{qian_large_2019,fu_berry_2023,zhang_large_2024}.   

We further performed excitation fluence-dependent measurements. The intensity of SHG emission from the CA film (red curve, left axis in Fig.~\ref{figure.3}\textbf{b}) increased quadratically as a function of excitation fluence $F_{\rm pump}$. In the small-fluence region, with $F_{\rm pump}$ values up to 0.12\,mJ/cm$^2$, the data exhibited a power-law dependence with an optimum exponent $k = 2.04\pm0.06$ when fit by the function $I_{\rm SHG} = A F_{\rm pump}^{k}$, as expected for SHG. However, as $F_{\rm pump}$ was further increased, a deviation from the quadratic dependence was observed. The deviation started at $F_{\rm pump} \sim 0.2$\,mJ/cm$^2$, which coincided with the onset of saturation absorption measured by power-dependent transmission (blue curve, right axis in Fig.~\ref{figure.3}\textbf{b}). Such saturation behavior is commonly seen in CNT assemblies~\cite{hussain_comparison_2019}, but the observed threshold fluence for the CA film was larger by one order of magnitude than what was previously reported in randomly oriented semiconducting CNT films containing multiple chiralities~\cite{wang_152_2016}. This demonstrates that an aligned, densely packed, and enantiomer-pure CNT film is a more robust optical material. 
The data shown in Figures 2 and 3\textbf{a} were acquired by keeping $F_{\rm pump}$ below 0.12\,mJ/cm$^2$ (marked by a black arrow in Fig.~\ref{figure.3}\textbf{b}) to stay in the quadratic regime.

\begin{figure*}[t]
    \centering
    \includegraphics[width=\linewidth]{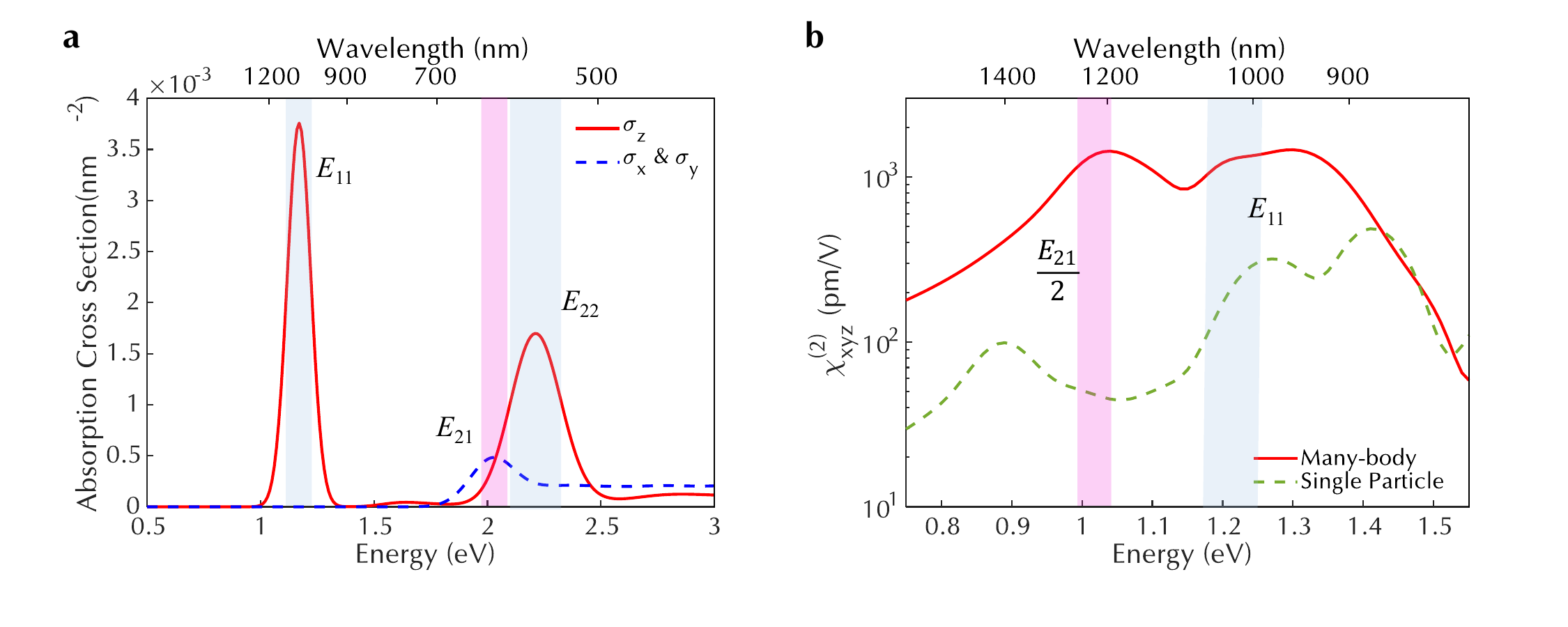}
    \caption{
        Many-body atomistic calculations reproducing the observed excitonic enhancement in (6,5)\textsuperscript{--} CNTs at the $E_{11}$ and $E_{12}$ resonances (blue and purple shaded regions). 
        \textbf{(a)} Linear absorption cross section for light polarized parallel ($\sigma_z$) and perpendicular ($\sigma_{x, y}$) to the nanotube axis. 
        \textbf{(b)} Nonlinear $\lvert\chi_{xyz}^{(2)}\rvert$ spectrum exhibits significant enhancement compared to the single-particle value using the same damping rate. The spurious peaks around 0.9~eV and 1.4~eV in the single-particle calculation are absent in the experimental results, highlighting the importance of many-body effects in nonlinear optical processes in CNTs.
    }
    \label{figure.4}
\end{figure*}

\subsection{Many-body Calculations of Exciton-enhanced Optical Nonlinearity}

Finally, we offer a comprehensive theoretical understanding of the observed excitonic enhancement of optical nonlinearity via many-body calculations (Supporting Information Section 8). We first calculated the linear absorption cross-section of (6,5)$^-$ CNTs via second-order perturbation theory following Ref.~\cite{JiUng2013}, with the excitonic transition dipole moments evaluated using the Bethe--Salpeter equation~\cite{Rohlfing2000}. As shown in Fig.~\ref{figure.4}\textbf{a}, symmetric excitonic absorption peaks are well-defined for light polarized parallel to the nanotube axis ($\sigma_{z}$), which is qualitatively different from previous theoretical calculations for CNT systems based on single-particle band structure (Figure S10). Introducing a Gaussian broadening parameter $\gamma(E) = 0.07\,\text{eV} + 0.08\,\text{eV} \cdot (E - E_{11})$, which is proportional to the energy and loss channels~\cite{Hertel2008}, reproduces the full-width-at-half-maximum (FWHM) of the $E_{11}$ and $E_{22}$ excitonic peaks in the polarized attenuance measurements (Fig.~\ref{figure.1}\textbf{f}). Then, following Ref.~\cite{Boyd2020}, we calculated \replaced{the second harmonic}{$\lvert\chi_{xyz}^{(2)}\rvert$} spectra with the given $\gamma$ and compared the results based on the excitonic picture with those based on the single-particle picture, \added{and converted single-tube susceptibility to $\lvert\chi_{xyz}^{(2)}\rvert$ assuming densely packed triangular lattice, which results in the same packing density as the experiment} (Fig.~\ref{figure.4}\textbf{b} and Supporting Information Section 8).

Without any free parameters, the calculated peak positions and values of $\lvert\chi_{xyz}^{(2)}\rvert$ from the excitonic picture agree with the experimental results well, in contrast to those from the single-particle calculations. The latter shows an artefact peak around 1.4\,eV due to contributions from higher-lying bands, a clear feature in all previous calculations but inconsistent with the experiment. Another artifact around 0.9\'eV is due to the incorrect energy of $E_{12}=E_{21}=(E_{11}+E_{22})/2$ whose degeneracy is lifted in the many-body picture by the exchange interaction. Previous reported $\lvert\chi_{xyz}^{(2)}\rvert$ values based on the single-particle picture accidentally had the right order of magnitude by using an ideal constant $\gamma = 0.05$\,eV for single CNT~\cite{guo_linear_2004,pedersen_systematic_2009}, as opposed to our calculation using realistic and experimentally obtained $\gamma$ values for densely packed thin films (Figure S11). Meanwhile, we note that the many-body calculation appears to overestimate the nonlinearity at $E_{12}$ and $E_{13}$ resonance (the feature around 1.3\'eV). The fact that we did not observe a larger feature at 1.3\'eV implies that the damping rate of the $E_{13}$ transition in CNT films is larger than what we estimated in the simple model of linear energy dependence. Indeed, the $E_{13}$ resonance was not distinguishable in the linear absorption spectrum in experiments, making it impossible to evaluate its actual $\gamma$. 

\section{Conclusions}

In summary, we discovered giant second-order optical nonlinearity in wafer-scale, enantiomer-pure, aligned, and densely packed chiral (6,5)$^-$ CNT thin films. The dependence of the SHG emission intensity on sample orientation and optical polarization demonstrates that the nonlinearity arises from intrinsic structural chirality rather than contributions such as defects and surfaces. Control experiments with randomly oriented and mixed-enantiomer thin films show that alignment and enantiomer purity are crucial for the observed chirality-induced SHG. Nonlinear spectroscopy reveals two excitonic resonances at pump photon energy of 1.21\,eV and 0.96\,eV, which correspond to the absorption resonance with $E_{11}$ exciton and emission resonance with $E_{12}$ exciton, respectively. The second-order nonlinear susceptibility tensor element $\chi_{xyz}^{(2)}$ reaches $1.6\times 10^3$\,pm/V, comparable to the highest values experimentally reported in the near-infrared, in quantitative agreement with our many-body atomistic calculations with no free parameters. Our discovery opens a route for applying chip-scale films of chiral CNTs to optoelectronics, as well as nonlinear and quantum photonics.
Our method can potentially be extended to fabricate wafer-scale, flexible, and CMOS-compatible thin films of enriched chiral CNTs with variable diameters, which, combined with field and dielectric tunability~\cite{chernikov_electrical_2015,lin_dielectric_2014}, may allow large optical nonlinearity across a broad spectrum from the UV to NIR range. 

\section{Methods}\label{methods}

\subsection{Preparation of Enantiomer-Pure Single-Chirality (6,5)$^-$ CNTs}

We used enantiomer-pure single-chirality (6,5)$^-$ CNTs (also known as (5,6) or (11,$-$5) or left-handed (6,5) CNTs), using the convention of Ref.~\cite{wei_determination_2017}; note that Refs.~\cite{ghosh_advanced_2010,ao_differentiating_2016} used the opposite sign convention. In general, ($n$,$m$) CNTs are right-handed (left-handed) when $n>m$ ($n<m)$, and ($n+m$,$-m$) CNTs are the mirror image of ($n$,$m$) CNTs~\cite{PhysRevB.69.205402}. They were obtained by using gel chromatography employing mixed surfactant systems~\cite{yomogida_industrial-scale_2016,yomogida_automatic_2020,wei_high-yield_2018}. CoMoCAT SWCNTs (704148, Sigma-Aldrich) were dispersed in 1.0\% aqueous solution of sodium cholate (SC, \(>\)99\%, Sigma-Aldrich) using an ultrasonicator (Sonifier 250D, Branson) with 30\% output for three hours. After ultracentrifugation at 210,000 g for two hours, the top 90\% of the supernatant solution was collected and mixed with an aqueous solution of sodium dodecyl sulfate (SDS, $>$99.0\%, Sigma-Aldrich) to prepare 2.0\% SDS + 0.5\% SC solution. 

A two-step separation process using different surfactant concentrations~\cite{yomogida_industrial-scale_2016} was performed automatically at 19-20\,$^\circ$C using a high-performance liquid chromatography system (NGC chromatography system, BIO-RAD) with columns filled with gel beads (Sephacryl S-200 HR, GE Healthcare). First, the SWCNT solution was loaded into the first column equilibrated with 2.0\% SDS + 0.5\% SC solution, and non-adsorbed CNTs containing (6,5) CNTs were collected as filtrate. The filtrate solution is mixed with SC solution to prepare 0.5\% SDS + 0.5\% SC solution and used for the subsequent separation process. After loading the CNT solution into the second column equilibrated with 0.5\% SDS + 0.5\% SC solution, adsorbed CNTs were eluted by using mixed surfactant solution containing sodium lithocholate (LC, $>$98\%, Tokyo Chemical Industry Co., Ltd.), i.e., 0.5\% SDS + 0.5\% SC + $x$\% LC solution, where $x$ is the LC concentration increased stepwise~\cite{yomogida_automatic_2020}. 

The high-purity (6,5)$^-$ CNTs were collected by adding a gradient solution of 0.5\%SDS +0.5\%SC+0.029\% LC solution to 0.5\%SDS +0.5\%SC+0.030\% LC solution. For film preparation, the contained surfactants were replaced with 0.3\% sodium deoxycholate (DOC) by ultrafiltration (Amicon Ultra-15 with Ultracel-100 membrane, Merck Millipore). The obtained SWCNT solution was treated with ultrasonication with 30\% output for ten minutes and ultracentrifugation at 210,000g for one hour to remove aggregates and used for film preparation (CA and CR films). 

\subsection{CNT Thin Film Preparation}

For fabricating a CA film, the enantiomer-pure (6,5)$^-$ CNTs were aligned through the controlled vacuum filtration (CVF) method~\cite{he_wafer-scale_2016}. The mother sodium deoxycholate (DOC) suspended CNT solution was diluted apriory to filtration by pure water to reduce the concentration of DOC from 0.3\% to 0.075\%. A total volume of 4\,mL was filtered through the filter membrane (Whatman Nuclepore Track-Etched membranes, MilliporeSigma). The filtration occurs in three stages: the gravitational, pumping, and drying phases to maintain a high controllability over the nematic ordering of the nanotubes. \added{The environmental humidity was not controlled because the filtration process is known to be insensitive to humidity.}~\cite{doumani_engineering_2023,10298902,Saito_Controlled_2025,Zacheo_Efficient_2024,Imtiaz_Facile_2023,Doumani_cue} After filtration, the film was transferred onto a fused silica substrate (MTI corporation) by dissolving the polycarbonate membrane with chloroform, cleaning it with isopropanol (IPA), and blowing it with dry air at ambient temperature. 
For fabricating a CR film, instead of diluting the solution with pure water, 1.5\,mL of CNTs suspended in DOC with 0.3\% concentration was added into 8.5\,mL of diluted NaSCN (10\,mmol/L of Na$^+$). The suspension was then filtered out at a faster pace. The fast filtration and introduction of Na$^+$ into the solution help make the alignment more chaotic. 
For fabricating a RA film, the whole recipe can be found in our previous work~\cite{he_wafer-scale_2016}.

\subsection{Characterization of CNT Suspension and Aligned Films}

Sample morphology and thickness were characterized using a scanning electron microscope (FEI APREO) and an atomic force microscope (Park Systems NX20). Enantiometric purity (EP) was evaluated by recording the CD and attenuation spectra of the suspension using a CD spectrometer (Jasco 815). The uniaxially anisotropic refractive index was extracted from the Mueller matrix spectroscopic ellipsometry data acquired at three incident angles and three azimuthal orientations using the RC2 ellipsometer setup (JA Woollam Co.Inc).

\subsection*{Polarized Attenuance Spectra}
Polarized attenuance spectra within a $\upmu$m$^2$ scale optically resolvable area for three samples were acquired by a home-built optical transmission measurement setup. A condensed lens was used to focus the light beam from the lamp onto the sample. The reference (transmitted) signal was collected by the objective under illumination at the area without (with) the sample. A polarizer was placed between the condenser lens and the sample to control the incident light polarization. The signal was then guided into the spectrometer through an optical fiber to get a spectrum. 
After subtracting noises from the spectrometer, the transmission was acquired by calculating the ratio of the transmitted spectrum over the reference spectrum. Assuming negligible reflection, attenuation was acquired by $-\log(T)$, where $T$ is the intensity transmittance. By adjusting the magnification of the objective and the optical fiber's diameter, the probing area size can be controlled.

\subsection{Polarized Raman Spectroscopy}

To comprehensively analyze nematic order, we employ a state-of-the-art Raman spectroscopy technique. Operating at a pump wavelength of 532\,nm, we utilized a high-resolution Raman microscope equipped with a $50\times$ objective lens, delivering a focused beam with a spot size of 1.3 \,$\upmu$m$^2$. To discern the alignment of nanotubes within the nematic medium, we investigated three distinct polarized configurations, VV, VH, and HH, while probing the same spot in all three configurations. The VV configuration aligned the incident and scattered light polarizations parallel to the nanotube axis, effectively probing alignment parallel to the incident beam. In the VH configuration, the incident polarization remained parallel to the nanotube axis while the scattered polarization becomes perpendicular, allowing us to gauge alignment perpendicular to the incident beam. Finally, the HH configuration employed perpendicular polarizations for both incident and scattered light. Additionally, by incorporating polarized attenuance data at 532\,nm and using Equation~\ref{raman_eq}, we assessed the nematic order parameter in three dimensions, which is a reasonable approximation since the film thickness is much larger than the CNT diameter. With $\Delta = A_{\parallel}/A_{\perp}$ being the dichroic ratio~\cite{raman}, the nematic order parameter can be obtained as
\begin{equation}
\begin{split}
    S_{\rm Raman}=\frac{ 6\Delta I_{\rm VV}+3(1+\Delta)I_{\rm VH}-8I_{\rm HH}}{6\Delta I_{\rm VV}+12(1+\Delta)I_{\rm VH}+16I_{\rm HH}} \\ 
    \label{raman_eq} 
\end{split}
\end{equation} 

\subsection{Second Harmonic Generation Measurements}

Second harmonic generation (SHG) measurements were performed in a transmission geometry using the home-built setup, as shown in Figure S7. An idler beam generated from a TOPAS Prime Automated OPA (Spectra-Physics, USA) with a 5\,kHz repetition rate, a 100\,fs pulse duration, and a tunable wavelength from 1760\,nm--2100\,nm was used. A BBO nonlinear optical crystal was used to convert the laser wavelength to 880\,nm--1050\,nm via second harmonic generation, followed by a 1326\,nm short pass optical filter to block the fundamental laser beam. The laser then passed through a Berek polarization compensator (Model 5540, Newport Corp.) set in half-wave plate mode to make sure the laser polarization was corrected to be in the $yz$ ($y'z'$) plane in the laboratory (sample) coordinate system, as indicated in Fig.~\ref{figure.2}\textbf{a}. A convex lens with a focal length of 75\,mm was utilized to focus the laser beam onto the sample. The diameter of the laser spot was measured to be around 70\,$\upmu$m varying with the laser wavelength. The sample was also kept in the $yz$ plane and tilted along the $x$-axis in the laboratory coordinate system to generate chirality-induced SHG. The SHG signal was extracted by a 785\,nm short-pass filter and a 633\,nm short-pass filter. A polarizer was put right before the CMOS camera detector (PM001, Brigates Microelectronics, China) to analyze the signal's polarization. The detector recorded the SHG signal and by integrating the count from each pixel, subtracting the background, and correcting the quantum efficiency of the detector, the actual SHG photon count can be acquired. 
\added{The SHG system is calibrated via BBO crystal with known $\chi^{(2)}$ value. The BBO crystal orientation is carefully aligned in phase matching condition. By comparing the theoretical SHG intensity generated from BBO and the actual intensity measured by the detector (calculated by photon counts), the conversion rate between the measured signal of the system and the generated photon  on the wavelength of interest is around 0.65 counts per photon. The conversion is determined by multiple factors, including optical losses in the setup, the camera's quantum efficiency, and the camera's electronic gain. Then, the accuracy of the system calibration is verified by CVD-grown monolayer WS$_{2}$. The $\chi^{(2)}$ of the sample is measured to be ~0.6\,nm/V at 900\,nm excitation wavelength, agreeing with previous literature~\cite{bredillet_dispersion_2020}.}

\section{Acknowledgements}
R.X., J.L., and H.Z. acknowledge the U.S.\ National Science Foundation (DMR-2240106) and Robert A.\ Welch Foundation (C-2128). J.D., V.L., A.B., F.T., V.P., and J.K.\ acknowledge the U.S.\ National Science Foundation (PIRE-2230727). J.D., A.B., F.T., and J.K.\ also acknowledge support from the Robert A.\ Welch Foundation (C-1509), the Air Force Office of Scientific Research (FA9550-22-1-0382), and the Chan Zuckerberg Initiative (WU-21-357). Y.Y.\ and K.Y.\ acknowledge support from JSPS KAKENHI (JP20H02573, JP21H05017, JP22H05469 and JP23H00259), JST CREST (JPMJCR17I5), and US-JAPAN PIRE collaboration(JPJSJRP20221202). W.T.\ and J.L.\ acknowledge Robert A.\ Welch Foundation (C-1716). EB is supported by the Graduate Research Fellowship Program (GRFP) from U.S.\ National Science Foundation under grant number DGE-1842494. R.S.\ acknowledges support from JSPS KAKENHI (JP22H00283) and Yushan Fellow Program from Taiwan.

\section{Author Contributions}
A.B., J.K., and H.Z.\ conceived the project. Y.Y.\ and K.Y.\ synthesized the enantiomer-pure (6,5)$^-$ CNT suspension. J.D.\ fabricated the CNT film samples. K.Y., Y.Y., and J.D.\ performed circular dichroism spectroscopy. J.D.\ and N.H.\ measured the refractive index of the enantiomer-pure aligned CNT thin film. R.X., J.D., and N.H.\ measured the polarized attenuation spectra. J.D.\ and R.X.\ performed AFM measurements. J.D., W.T., and R.X.\ performed SEM measurements. J.D.\ performed polarized Raman measurements. R.S.\ examined issues within the theoretical framework of carbon nanotubes. R.X.\ performed SHG experiments and analyzed the data. V.L.\ conducted the many-body atomistic calculations under the supervision of V.P. R.X., J.D., A.S., M.T.,E.B., J.L.,F.T., V.P., A.B., and H.Z.\ wrote the manuscript with input from all authors. All authors discussed the data.

\section*{Competing Interests}
The authors declare no competing financial interests.  

\section{Supporting Information}
The Supporting Information is available free of charge online.

Characterization of the Purity of CNT Chirality and the Purity of the Enantiomer (Figure S1); Measurement of the Film Thickness (Figure S2); Electron Microscopy of Random and Aligned Films (Figure S3); Chip-Scale- and micro-Polarized Absorbance and Raman Spectroscopy of Aligned (6,5)$^-$ nanotubes (Figure S4, Table S1); Polarized Attenuation Spectra for Random \& Racemic CNT Films (Figure S5); Refractive Indices of CNT Thin Film Sample (Figure S6); Experimental Measurement of Second-order Nonlinear Optical Susceptibility of (6,5)$^-$ CNT Thin Film (Figure S7, S8); Theoretical Calculations for Second-order Nonlinear Optical Susceptibility of (6,5)$^-$ CNT (S9--S11).

\section{Associated Content}
Xu R; Doumani J; Labuntsov V; Hong N; Samaha A-C; Tu W; Tay F; Blackert E; Luo J; Tahchi ME; Gao W; Lou J; Yomogida Y; Yanagi K; Saito R; Perebeinos V; Baydin A; Kono J; Zhu H. Chip-Scale Aligned Chiral Carbon Nanotubes Exhibiting
Giant Second Harmonic Generation. 2026, 2407.04514. ArXiv. https://doi.org/10.48550/arXiv.2407.04514 (accessed 04/28/2026)

\section{Data Availability}
The data underlying this study are available in the published article and its Supporting Information.

\bibliography{references}

\end{document}